\documentclass[9pt,twocolumn]{article}
\usepackage[a4paper,margin=1.8cm]{geometry}
\usepackage{multicol}
\usepackage{amsmath,amssymb}
\usepackage{graphicx}
\usepackage{xcolor}
\usepackage{gensymb}
\usepackage{lineno}
\usepackage{authblk}

\title{Event-based Speckle Interrogation for High-Bandwidth Multi-point Optical Fiber Sensing}

\author[1,2]{Tomás Lopes\thanks{tomas.j.lopes@inesctec.pt}}
\author[1,2]{Joana Teixeira}
\author[1,2]{Vicente V. Rocha}
\author[1]{Tiago D. Ferreira}
\author[1]{Catarina S. Monteiro}
\author[1,2]{Pedro A. S. Jorge}
\author[1,2]{Nuno A. Silva}

\affil[1]{Centre for Applied Photonics, INESC TEC, Rua do Campo Alegre 687, 4169-007 Porto, Portugal}
\affil[2]{Department of Physics and Astronomy, Faculty of Sciences, University of Porto, Rua do Campo Alegre s/n, 4169-007 Porto, Portugal}

\affil[ ]{\textit{These authors contributed equally to this work.}}

\begin{document}
\maketitle

\begin{abstract}
    Speckle-based fiber optic sensors are well-known to offer high sensitivity but are strongly limited on the interrogation side by low camera frame rates and dynamic range. To address this limitation, we present a novel interrogation framework that explores event-based vision to achieve high throughput, high bandwidth, and low-latency speckle analysis of a multimode optical fiber sensor. In addition, leveraging a tensor-based decomposition of the raw event streams through multi-point calibration and machine-learning optimization, our approach also proves capable of isolating simultaneous deformations applied at distinct points. 
The experimental results validate the methodology by separating the signals of four piezoelectric actuators over a 400\,Hz –20\,kHz range with minimal crosstalk applied over varying distances from 3\,cm to 75\,cm. Finally, extending the impact of the work with an acoustic sensing proof-of-concept, we have coupled the fiber to two plastic enclosures and recovered separable audio signals between 400 and 1.8 kHz with minimal waveform distortion. Overall, these results establish event-driven speckle interrogation as a new versatile platform for real-time, multi-point acoustic sensing and pave for its application in complex and unstructured environments in future works.
\end{abstract}



\section{Introduction}\label{sec1}

The propagation of coherent light through complex optical media, such as multimode fibers, typically results in the generation of a complex intensity pattern, referred to as speckle\cite{rotter2017light}. This behavior is caused by the interference of the various spatial modes and is often considered detrimental for optical communications and imaging applications due to its sensitivity to environmental changes, e.g., strain, temperature, and pressure \cite{hsu2017correlation,byrnes2020universal}. However, in the absence of significant loss and non-linear effects, the underlying linear and deterministic nature of light propagation in waveguides ensures that the system can be fully characterized mathematically by a transmission matrix, mapping a set of input modes to corresponding output modes \cite{popoff2010measuring}. Fueled by the emergent development of wavefront-shaping techniques and holography \cite{gigan2022roadmap}, this formalism led to a paradigm shift in the characterization of complex media, transforming the speckle into a controllable resource that now opens opportunities beyond traditional applications, including the field of distributed sensing with optical fibers. 

In conceptual terms, fiber-based speckle sensing leverages the intrinsic high-dimensional encoding provided by the interference pattern, also called a specklegram. Indeed, the high-dimensional characteristics of the output signal have been demonstrated to enable enhanced sensitivity \cite{gutierrez2024reaching}, reduced noise levels via statistical methods \cite{bouchet2021maximum}, and even support distributed sensing\cite{murray2019speckle,cuevas2018machine,gao2023spatially}. Building on foundational works on the topic \cite{spillman1989statistical, wu1991sensing, okamoto1988multimode}, multiple architectures and solutions have emerged, including transmission schemes that treat the whole specklegram as a point sensor for temperature, strain, and pressure \cite{wang2017speckle, reja2024multimode}, differential and reference-less designs for dynamic sensing \cite{spillman1989statistical, yu1993submicrometer, gupta2008qualifying, lujo2008fiber, fujiwara2022optical}, reflective configurations with pulsed sources for location \cite{murray2019speckle}, and multi-section and concatenated solutions with distinct types of fibers to exploit modal diversity \cite{fujiwara2019optical, efendioglu2017review}. More recently, and fueled by the emergence of machine learning, fiber speckle sensors have also been proposed for multi-point and distributed sensing, exploiting multivariate regression patterns \cite{wei2021neural, gao2023spatially,cuevas2018machine}. 

Yet, these capabilities are associated with the complexity of the intensity pattern generated at the output, meaning that the interrogation procedure requires a multi-pixel detector (e.g., multiple detectors or a camera sensor), which strongly bounds sensing performance. On one hand, it restricts operating frequency bandwidth (typically 10-100\,Hz due to the Nyquist theorem for conventional cameras) and dynamic range (regions of the speckle may be saturated, whereas others may not). On the other hand, data-driven methodologies require intricate and computationally-heavy calibration procedures\cite{gao2023spatially,cuevas2018machine}, hardly suitable for real-world applications. 

In the context of these challenges and in theory, the interrogation of these sensors with neuromorphic event-vision sensors (EVS) may appear particularly promising. Unlike conventional imaging sensors, an EVS camera does not record absolute intensity values but instead detects changes in intensity that exceed a predefined threshold at each pixel (with a given polarity associated with the signal of the variation)\cite{lenero2018applications}. This paradigm shift brings three key benefits for speckle sensing. First, as events are triggered only by actual speckle dynamics, the effective bandwidth extends into the MHz regime and sub-microsecond latencies become achievable, thus vastly outpacing typical CMOS cameras \cite{lenero2018applications}. In addition, the logarithmic response of each pixel confers a much larger dynamic range(up to 120dB), preventing saturation of bright grains while still capturing low‐intensity fluctuations without underflow \cite{gallego2020event}. Finally, due to its asynchronous operation, the data rate scales with the rate of physical perturbation rather than with the megapixel count, reducing computational load and enabling real‐time, on‐edge multivariate decoding \cite{gallego2020event}.

Leveraging on this opportunity, this manuscript explores the use of neuromorphic event-based vision sensors (EVS) to deploy a multi-point vibration and acoustic sensing solution based on a fiber-optic speckle sensor. First, based on transmission matrix formalism for multimode fibers\cite{ploschner2015seeing, popoff2010measuring, bianchi2012multi}, we develop a tensor-based decomposition of the raw event streams and design a multi-point calibration and machine-learning optimization to isolate simultaneous deformations at distinct fiber locations. Then, experimentally, we first characterize the frequency response of the system, before demonstrating the accurate separation of four simultaneous signals applied to individual piezoelectric actuators across a 400 Hz–20 kHz band over distances from 3 cm to 75 cm. Finally, extending the proof-of-concept to acoustic sensing, we also recover separable yet simultaneous audio signals between 400 Hz and 1.8 kHz with negligible distortion. The framework and results establish event-driven speckle interrogation as a versatile, high-throughput framework for real-time, multi-point fiber-optic sensing in challenging, unstructured environments, and possibly competing with conventional distributed acoustic sensing (DAS) - e.g., Rayleigh or Brillouin - technologies in the near future.


\section{Theoretical Model}\label{sec2}


\begin{figure*}
\centering
\includegraphics[width=\linewidth]{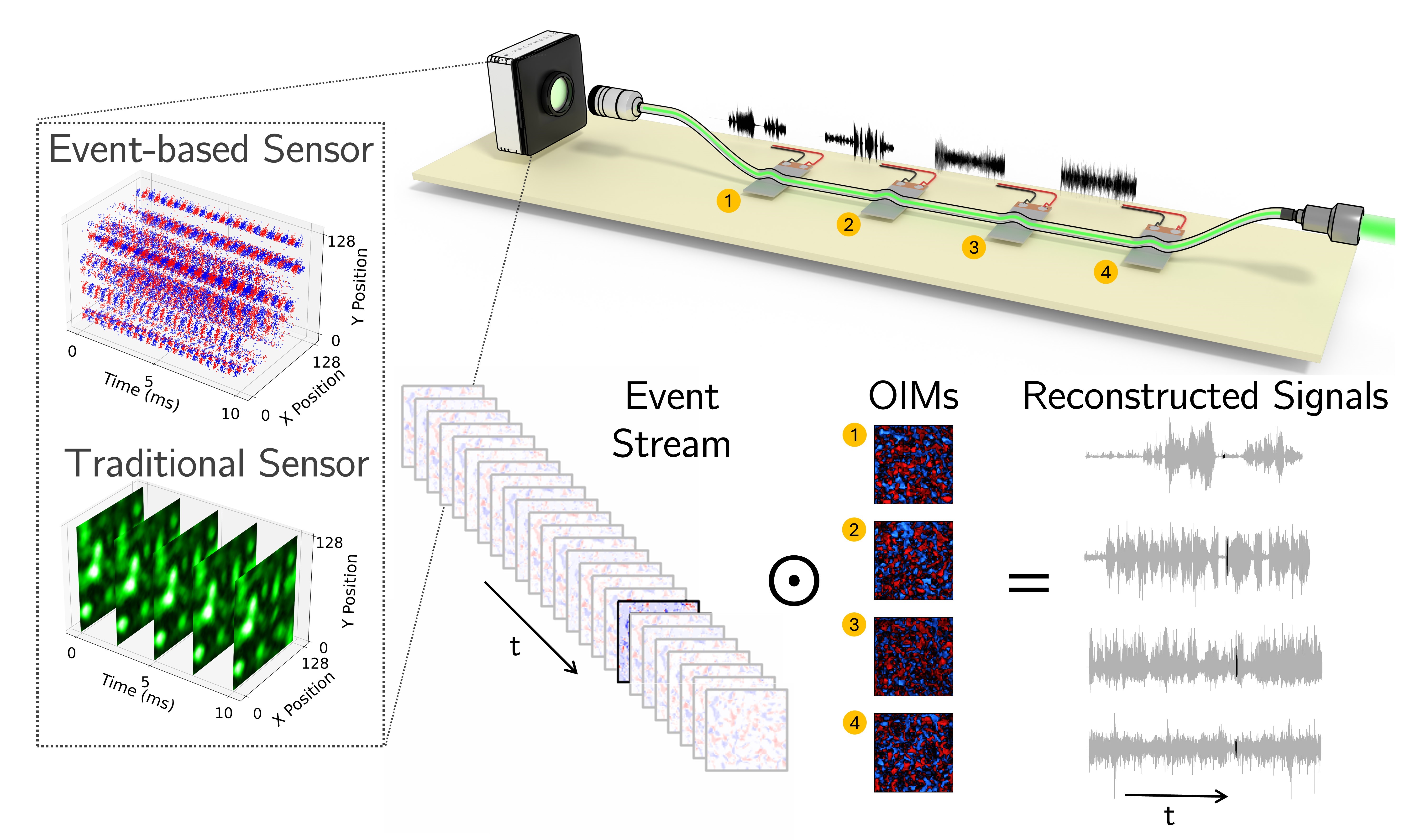}
\caption{Event-based sensing for high-speed interrogation of speckle patterns in a multimode optical fiber. Vibrations from multiple piezoelectric membranes induce dynamic speckle variations captured by an event-based vision sensor. A PyTorch model optimizes individual speckle pattern-inspired tensors, which represent the optimal interrogation modes (OIM), to separate the simultaneous deformation signals.}
\label{fig:general_image}
\end{figure*}

The propagation of optical fields in multimode optical fibers occurs via multiple guided modes, each characterized by distinct spatial distributions and propagation constants. When coherent light exits such media, the resulting output field distribution represents the superposition of all excited modes, creating a complex intensity pattern at the observation plane known as a speckle pattern. Any external perturbation, such as mechanical deformation or temperature fluctuation, induces significant changes in this pattern. These changes arise from two primary mechanisms: (1) inter-modal coupling, where energy transfers between previously orthogonal modes, and (2) differential phase accumulation, where each mode experiences unique phase shifts due to its different propagation constant. Together, these effects transform the output speckle pattern in a manner that appears random but is, in fact, deterministic. Indeed, using the decomposition of the intensity into modes
\begin{equation}
I(\mathbf{r},t) = \left|\sum_{j} a_j(t) \psi_j(\mathbf{r}) \exp(i\phi_j(t))\right|^2
\end{equation}
where $a_j(t)$ and $\phi_j(t)$ are the time-dependent amplitude and phase of mode $j$, and $\psi_j(\mathbf{r})$ represents the spatial field distribution of each mode. Under mechanical perturbations, the relative intensity change can be approximated as
\begin{equation}
\frac{\Delta I}{I_0} \approx \frac{2}{I_0}\sum_{j, k} a_j a_k \cos(\phi_j - \phi_k) \Delta\phi_{jk}
\end{equation}
where $\Delta\phi_{jk}$ represents the phase changes between modes induced by the perturbation, thus encoding detailed information about the deformation in a high-dimensional optical response \cite{gutierrez2024reaching}. While the complete mathematical description of light propagation under such perturbations becomes highly complex due to spatially-dependent coupling coefficients and even nonlinear elasto-optic effects \cite{tiwari1999nonlinear}, the transmission matrix formalism still offers a powerful framework to characterize such complex media. Indeed, assuming a small and localized perturbation, the response of the optical field at the output of a multimode fiber can be effectively approximated by
\begin{equation}
     \boldsymbol{I}(x) = | \bar{\boldsymbol{M}}_2 \bar{\boldsymbol{D}}(x;\delta)  \bar{\boldsymbol{M}}_1 \ \boldsymbol{E}_{\text{in}}|^2,
     \label{eq:intensity_under_deformation}
\end{equation}
where $\bar{\boldsymbol{M}}_{1,2}$ represent the transmission matrices governing light propagation through the fiber segments before and after the perturbation point, respectively, while $\bar{\boldsymbol{D}}(x;\delta)$ characterizes the perturbation-induced modification of the transmission matrix at position $x$ by an applied stimulus $\delta$.

Assuming a linear regime for small perturbations, one can linearize the transmission matrix $\bar{\boldsymbol{{D}}}(x)$ in first order as
\begin{equation}
    \bar{\boldsymbol{D}}(x) \approx \bar{\boldsymbol{D}}_0 + ( \partial_{\delta}  \bar{\boldsymbol{D}} ) \delta + O(\delta^2),
\end{equation}
being $ \bar{\boldsymbol{D}}_0$ the unperturbed transmission matrix. Expanding equation \ref{eq:intensity_under_deformation} as demonstrated in Section 2 of the Supplementary Information, yields the intensity in the form
\begin{equation}
    \boldsymbol{I}(x;\delta) \approx \boldsymbol{I}_0 + \partial_\delta \boldsymbol{I}\delta  + O(\delta^2) , 
\end{equation}
which directly follows the definition of the first derivative. In physical terms, the intensity pattern at the output can be described as a linear evolution of the pattern at a previous point, provided the applied deformation $\delta$ is small enough to be considered within the linear regime perturbation coordinate. 

Generalizing the result to a set $\{\delta_i\}$ of $N$ simultaneous deformations applied at $\{x_i\}$ distinct points in the optical fiber, the linearized expansion generalizes to
\begin{equation}
    \boldsymbol{I}(\{x_i\}; \{\delta_i\}) \approx \boldsymbol{I}_0 + \sum_{i = 1}^{N} (\partial_{\delta_i} \boldsymbol{I}) \delta_i.
    \label{eq:separation}
\end{equation}

By rearranging terms and considering that $\boldsymbol{I}$ is a vector $I(x) \in \mathbb{R}^{m}$ with $m$ representing the number of features extracted from the intensity pattern, i.e., the number of macro-pixels in the detection scheme. The deformation profile within the optical fiber can be reconstructed from local intensity variations:
\begin{equation}
\begin{aligned}
    \begin{bmatrix}
        \delta_1\\
        \vdots\\
        \delta_N
    \end{bmatrix} &\approx 
    \bigl[\boldsymbol{I}(\delta_1, ..., \delta_N) - \boldsymbol{I}_0\bigr] \cdot \frac{ (\partial_\delta \boldsymbol{I})^+ }
    {\bigl\lVert (\partial_\delta \boldsymbol{I})^+ \bigr\rVert^2}.
\end{aligned}
\label{eq:def}
\end{equation}
where the $^+$ symbol denotes the Moore-Penrose pseudo-inverse and 
\begin{equation}
    \partial_\delta \boldsymbol{I} \equiv
    \begin{bmatrix}
        \partial_{\delta_1}\,\boldsymbol{I}\\
        \vdots\\
        \partial_{\delta_N}\,\boldsymbol{I}
    \end{bmatrix}
\end{equation}
being each of the terms a formal partial derivative
\begin{equation}
\partial_{\delta_i} \boldsymbol{I}\ = lim_{\delta_i\rightarrow0} \frac{\boldsymbol{I}(0,...,\delta_i,...,0) - \boldsymbol{I}_0}{\delta_i}
\end{equation}
Put in this form, it is clear to establish a relation between the capability of a fiber speckle sensor to spatially resolve the applied stimulus and the rank of the matrix $\partial_\delta \boldsymbol{I}$ (which ultimately relates to the number of spatial modes). Indeed, each column of the inverse matrix acts as an interrogation mode to be applied to the measured signal: spatial weight that determines the contribution of each macro-pixel to the overall signal and a corresponding temporal signature for time-varying $\delta_i$. 

However, in practice, determining the adjoint matrix $\partial_\delta \boldsymbol{I}^{+}$ is an ill-posed problem and thus very sensitive to noise, which may, in theory, lead to sub-optimal results. A way to circumvent this limitation is to determine each optimal interrogation mode $\boldsymbol{O}_i$ using a data-driven approach such that $\boldsymbol{O}_i \approx \partial_{\delta_i} \boldsymbol{I}$. For this, in this work, we will optimize a feature-wise linear layer model, spanning across $N$ deformations, which is optimized using backward propagation to minimize the loss function
\begin{equation}  
\mathcal{L} = \sum_{i=1}^{n} \frac{\text{Var}(\delta^R_i(t) \, | \, \delta_i(t) = 0) + \beta}{\text{Var}(\delta^R_i(t) \, | \, \delta_i(t) \neq 0)},
\label{eq:loss}
\end{equation}
where $\delta^R_i(t) = \bigl[\boldsymbol{I}(\delta_1, ..., \delta_N) - \boldsymbol{I}_0\bigr]\odot\boldsymbol{O}_i$ is the reconstructed signal and $\beta$ is a physical parameter proportional to the inverse density of the pattern, used to control the uniformity of the optimized tensor weights. This approach is schematically represented in Figure \ref{fig:general_image}.


\subsection{Event-Based Vision Interrogation}

When it concerns interrogation of the speckle pattern, i.e., measuring $\boldsymbol{I}$, standard CCD and CMOS sensors present several challenges. First, we have the limited frame-rate of the sensor itself (typically in the range of hundreds of frames per second, at most), which translates into a limited bandwidth (from Hz to few kHz) that strongly limits the span of applications of the speckle-based sensors in acoustic and vibration sensing. Then, these sensors have a limited dynamic range, meaning that under- or over-exposed regions of the speckle may not contribute to usable signal, which strongly bounds the performance in this regard. Finally, these sensors record instantaneous intensity over a finite integration period, meaning that for dynamic sensing requiring differential schemes\cite{cuevas2018machine}, a significant post-processing overhead is necessary, adding computing time as well as higher energy consumption.

To circumvent the limitations of these sensors in dynamic scenes, event-based vision sensors (EVS) (also known as neuromorphic cameras) have recently emerged as an alternative. In conceptual terms, an EVS mimics biological vision systems by detecting only changes in intensity that exceed a predefined threshold at each pixel, thus providing native differential detection. In particular, in the context of speckle interrogation, this comes with various advantages, from low latency (asynchronous stream, meaning that computational load scales with speckle dynamics rather than pixel count), to high-speed interrogation (up to microsecond timescale), and large dynamic range (preventing saturation under bright illumination). As for each captured event, the recorded variation follows
\begin{equation}
    \Delta \boldsymbol{I} = \text{sgn} \left( \bigl[\boldsymbol{I}(x) - \boldsymbol{I}(x - \Delta x)\bigr] \right),
\end{equation}
where $\text{sgn(a)}$ is the standard \textit{signum} function. By accumulating events over a chosen integration time, one recovers a quantity proportional to the true intensity change, effectively reconstructing the analogue $\Delta \boldsymbol{I}$ signal which natively provides data proportional to the variation $[\boldsymbol{I}(\delta_i)-\boldsymbol{I}_0]$ of the speckle pattern intensity with respect to stimulus $\delta_i$ applied during the integrated period of time. This framework implies that while exact deformation amplitudes cannot be reconstructed from the binary event data alone, the system remains highly effective for capturing temporal signal dynamics with unprecedented time resolution. 




\section{Results and Discussion}

\subsection{Experimental Setup}
Our experimental setup consists of a 532\,nm laser (CNI MSL-DS-532) whose beam is coupled into a 4-meter-long step-index multimode optical fiber (50\,µm core, 125\,µm cladding, 0.2\,NA), with small sections (2–3\,cm) placed in direct contact with four piezoelectric membranes (PHUA3015-049B-00-000, TDK). The first two membranes are placed approximately 75\,cm apart, while the distance between the second and third membranes is only 3\,cm. The last membrane is positioned 105\,cm apart from the previous one. To drive the piezoelectric membranes, a NI 6259 DAQ supplies voltage signals between -10 and 10\,V. A metal disk weighing 250\,g is placed on top of each piezoelectric membrane to warrant contact with the optical fiber.

At the output end of the optical fiber, the resulting speckle pattern is magnified using a $10\times$ microscope objective. This magnified speckle pattern is subsequently recorded by an event-based camera (EVK4, Prophesee), enabling high-speed and asynchronous imaging suitable for capturing dynamic variations in the optical signal.


\subsection{Determination of the Optimal Interrogation Modes for Multi-point Localization}
\label{sec:calib}


The theoretical framework presented in equation \ref{eq:separation} suggests that with perfect knowledge of the response of the system, complete separability of signals from different deformation points should be achievable. However, several practical considerations significantly complicate this idealized scenario. In real-world implementations, we observed substantial cross-talk between signals due to various factors. Modal overlap occurs as different deformation locations can excite partially overlapping sets of optical modes, creating similar signatures in certain regions of the speckle pattern. Measurement noise from the binary nature of event detection introduces quantization effects that impact reconstruction accuracy. Environmental factors such as intrinsic thermal fluctuations and ambient mechanical vibrations introduce background noise that affects signal fidelity \cite{zhang2011photonic, zhang2006fiber}. Additionally, mathematical limitations arise as the Moore-Penrose pseudo-inverse used in our reconstruction algorithm inherently produces smoothed solutions that distribute signal energy across multiple spatial locations.
Initial experiments applying the direct theoretical approach revealed significant cross-talk between signals originating from different fiber locations, limiting the ability of the system to accurately localize and quantify multiple simultaneous perturbations.

To address these limitations, we developed a tensor-based optimization approach that substantially improves signal reconstruction fidelity. Rather than relying solely on the theoretical model, our data-driven extraction approach (detailed in section \ref{sec2}) incorporates calibration measurements to learn the actual response characteristics of the system. While this approach leverages the underlying linear relationship between small deformations and corresponding intensity changes, it may still account for the practical deviations of the experimental system from this regime. The optimization process effectively creates a set of weight masks that maximizes sensitivity to perturbations at their assigned locations while minimizing cross-talk from other places.

For determining the OIM under the data-driven framework, we applied controlled perturbations using piezoelectric actuators positioned at multiple discrete points along the optical fiber. These actuators were driven with sinusoidal signals of known frequency ($f=1000$\,Hz) and amplitude, creating reference deformations well-defined in the spectral domain. The event-based camera inherently captures intensity changes rather than absolute values, which directly provides information proportional to the derivatives of speckle pattern intensity with respect to applied deformations. Nevertheless, to increase signal-to-noise ratio and average the effects of the binary signal, for practical signal reconstruction we accumulate detected events over intervals of $\Delta T =  25\mu s$, which, according to Nyquist's Theorem, enables the detection of signals across the full audible spectrum (up to 20\,kHz) while maintaining sufficient signal-to-noise ratio. The recorded event streams were then utilized to reconstruct a signal by element-wise multiplication according to equation $\delta^R_i(t) = \bigl[\boldsymbol{I}(\delta_1, ..., \delta_N) - \boldsymbol{I}_0\bigr]\odot\boldsymbol{O}_i$, and the optimal interrogation modes $\boldsymbol{O}_i$ were optimized as described in section \ref{sec2} using a Pytorch implementation.


\begin{figure}[h!]
    \centering
    \includegraphics[width=\linewidth]{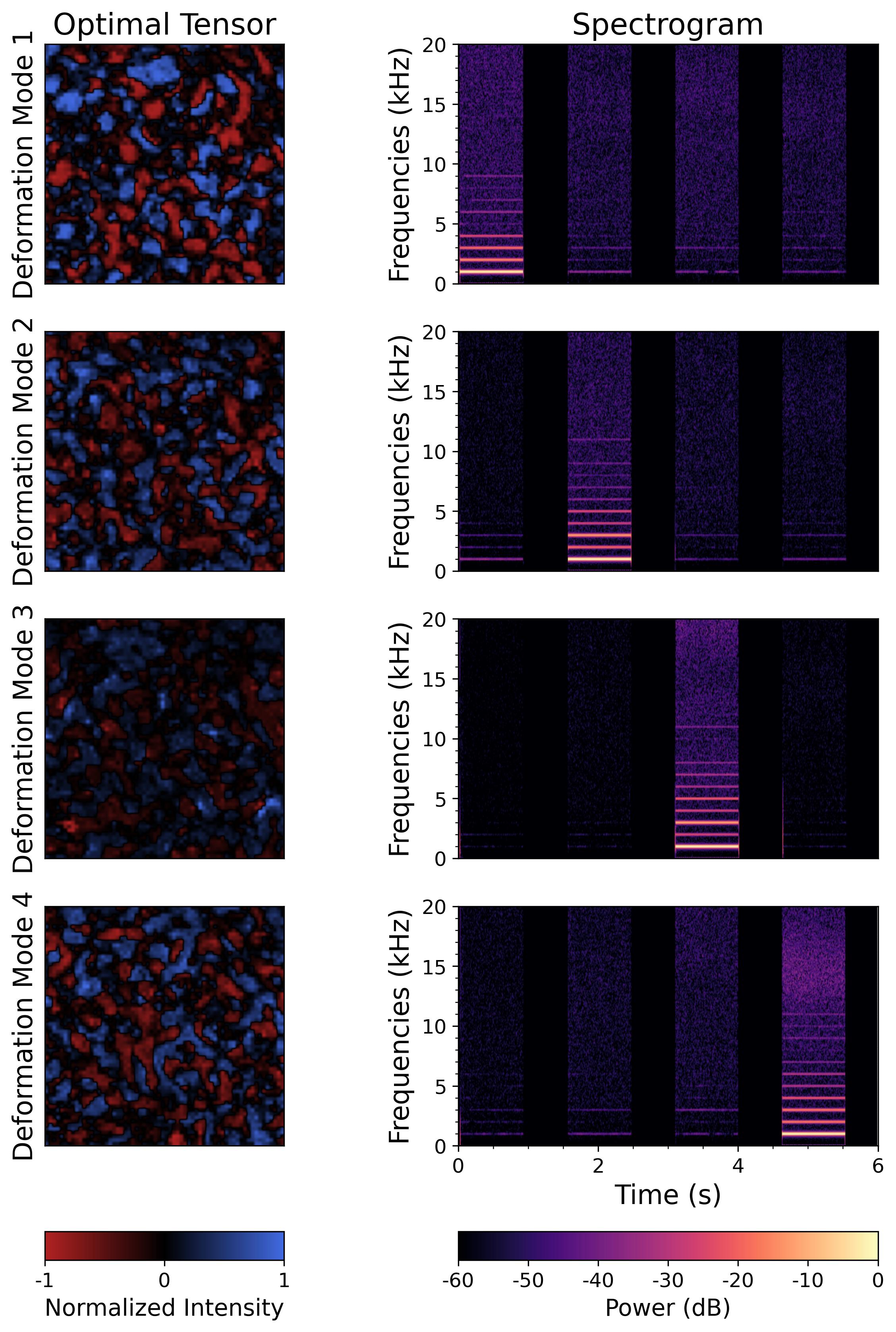}
    \caption{Optimal interrogation modes for deformation recovery and respective spectrograms for calibration, along with the corresponding spectrograms displaying the separability of the signals, demonstrating its distributed capabilities. To simplify their representation, the individual tensors were normalized to their maximum value. }
    \label{fig:calibration_distributed}
\end{figure}

Figure \ref{fig:calibration_distributed} demonstrates the effectiveness of this approach by presenting the optimized tensors for each deformation signal $\delta^R_i(t)$, alongside corresponding spectrograms for $f=1000$\,Hz sinusoidal signals applied sequentially to different piezoelectric membranes and interrogated with distinct optimal interrogation modes. As can be seen qualitatively from the spectrograms of each reconstructed signal, the fundamental vibration frequency for each deformation location is isolated and stronger at $f=1000$\,Hz, along with several harmonics that arise from minor nonlinearities in the piezoelectric actuators and the optical response.

Note, however, that while the optimization process significantly reduces cross-talk between signals, some residual interference remains visible in the spectrograms. Importantly, this residual cross-talk appears to depend primarily on the optimization parameters rather than on the physical position or amplitude of the deformation. This observation suggests that further refinement of the optimization model could yield additional improvements in signal separation, potentially approaching the theoretical limit of completely independent sensing zones along a single multimode fiber. 


\subsection{Frequency Response}

As previously referred, one of the key advantages of using the EVS is the possibility of increasing the bandwidth on the interrogation side compared with conventional camera interrogation. To characterize the capabilities of our system in this regard, we proceed by evaluating its performance across a broad range of frequencies, applying a linear frequency sweep (chirp) to assess the frequency-dependent sensitivity of the system systematically.

\begin{figure}[h!]
    \centering
    \includegraphics[width=\linewidth]{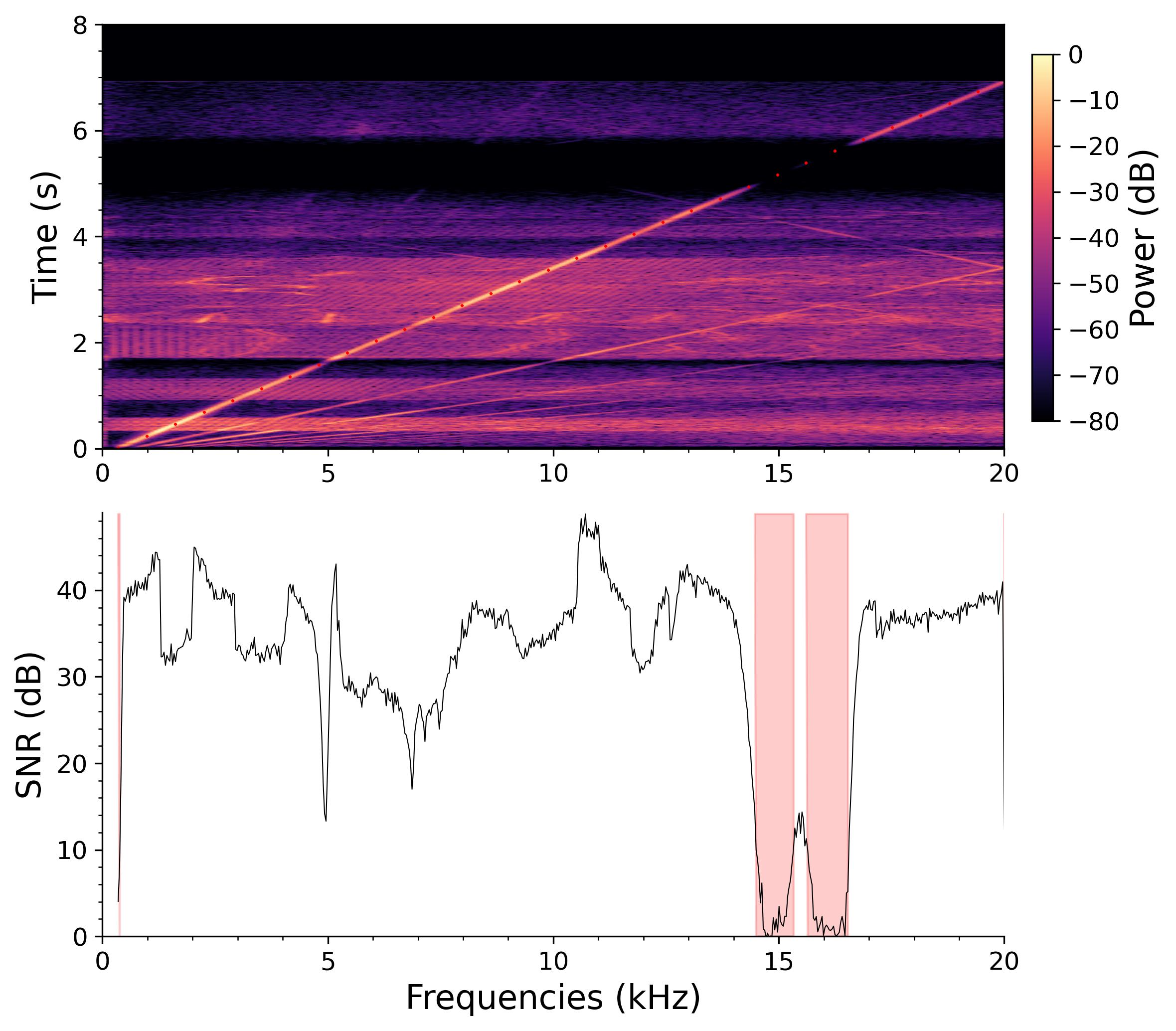}
    \caption{Spectrogram obtained from the reconstruction of a sinusoidal frequency ramp applied to one of the piezo membranes (top), where the dotted red line follows the theoretical placement of the ramp. The frequency response of the sensor corresponds to the signal-to-noise ratio between the points in the frequency ramp and the rest of the spectrogram (bottom plot). The frequency range covered in this test is between 350 - 20\,000 Hz. The regions marked in red correspond to regions where the signal-to-noise ratio is lower than 10\,dB. }
    \label{fig:results_distributed}
\end{figure}

After the determination of the OIM, a linear frequency ramp ranging from 350\,Hz to 20,000\,Hz was applied to one of the piezoelectric actuators, allowing us to continuously probe the response of the system across the entire audible spectrum. Figure \ref{fig:results_distributed} presents the resulting normalized spectrogram of the reconstructed signal. To quantitatively assess performance across frequencies, we calculated the power-to-noise ratio for each frequency component in the ramp. This was determined by computing the difference between the maximum power in each frequency column of the spectrogram and the average background power along that same column, expressed in decibels, as
\begin{equation}
    \mathrm{SNR} ({f_\text{target}})_{dB} = \mathrm{S}({f_\text{target}}) - \sum_{f\neq f_\text{target}}^{f_N} \frac{\mathrm{S}({f})}{N-1}
\end{equation}

where $f_\text{target}$ is the frequency for which the response of the sensor is being evaluated, $\mathrm{S}({f_\text{target}})$ is the corresponding power (in decibels), obtained from the spectrogram, and N is the total number of frequencies to evaluate.

It is important to note that both the spectrogram visualization and SNR calculations were performed after normalizing the data to its maximum value, ensuring consistent comparison across the frequency range. Also, in order to isolate the optical sensing performance from piezoelectric actuator characteristics, the frequency response data presented has been normalized by the frequency-dependent deformation amplitude produced by the piezoelectric membrane. This normalization procedure, detailed in the supplementary material, compensates for the non-flat frequency response of the actuators themselves, revealing the true sensitivity profile of the optical fiber sensing system.

The results obtained demonstrate robust signal reconstruction capabilities up to 20\,kHz, confirming the suitability of the system for acoustic and vibration sensing across the full human audible range. However, we observed a notable reduction in signal-to-noise ratio around $f=15$\,kHz. This localized sensitivity reduction likely stems from mechanical coupling effects between the piezoelectric actuator and the optical fiber, which can vary with the positioning of the weights used to maintain fiber tension. The precise frequency response depends on the specific mechanical configuration at each deformation point, highlighting the importance of stable mounting for consistent operation.

\subsection{Complex Signal Location and Reconstruction}
\label{sec:complex_audio}

\begin{figure*}[h!]
    \centering
    \includegraphics[width=\linewidth]{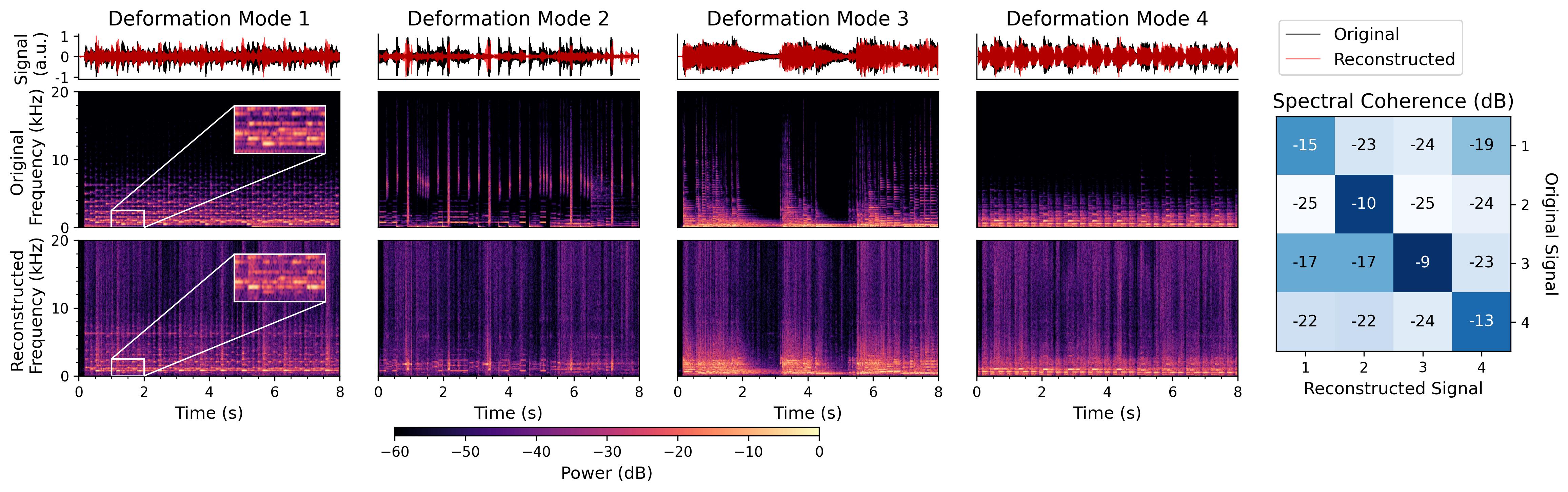}
    \caption{Original and reconstructed signals for the four deformation modes, as well as the corresponding spectrograms. The frequencies applied to the piezoelectric membranes correspond to four royalty-free music clips. A spectral coherence index confusion matrix is presented as a metric for the spectral similarity between the original and reconstructed signals, as well as the crosstalk between different deformation modes. }
    \label{fig:music}
\end{figure*}

To further evaluate the separability and fidelity of our signal reconstruction approach, we extended our analysis to more complex audio signals. In this experiment, we simultaneously applied different sound type selections to multiple piezoelectric actuators positioned along the fiber. We deliberately selected music from diverse genres to ensure the representation of varied spectral content, from low-frequency bass components to high-frequency elements and vocals.
For signal reconstruction, we applied the same optimal interrogation modes extracted during the calibration process described in section \ref{sec:model}. The optimized projection matrices were directly multiplied by the accumulated event frames without additional signal processing. Figure \ref{fig:music} presents the spectrograms of the reconstructed audio signals for each sensing location.

The results demonstrate strong signal separability and reconstruction fidelity. By visual analysis, it is possible to conclude that each reconstructed audio stream preserves the distinctive spectral and temporal characteristics of its corresponding musical input, including fundamental frequencies, harmonic structures, and temporal dynamics. Particularly notable is the minimal cross-talk between channels, even when reconstructing signals with overlapping frequency content. This confirms that the event patterns generated by deformations at different fiber locations remain sufficiently distinct to enable independent signal recovery. The audio files resulting from these reconstructions can be accessed \cite{Lopes_Harnessing_Disorder_Tensor-Driven}. To quantitatively characterize this behavior, we performed a weighted spectral coherence analysis (WSC), calculated as the weighted average of spectral coherence over the power spectral density of the original signals, presented in decibels, using the formula
\begin{equation}
    \mathrm{WSC}_{dB}
        =10 \log_{10}\!\left(
        \frac{\sum_f C_{ij}(f)\,P_{ii}(f)}
        {\sum_f P_{ii}(f)}
    \right),
\end{equation}
where $i$ and $j$ represent the reference and reconstructed signals, respectively, and $C_{ij}$ represents the standard magnitude squared spectral coherence per frequency bin
The diagonal entries in this matrix confirm the high level of spectral similarity between the original and reconstructed audio files, verifying the preservation of the key frequency components for each deformation mode. Moreover, the comparison between non-corresponding signal pairs reveals reduced cross-talk across deformation modes, as evidenced by lower off-diagonal spectral coherence values. While these values are not minimal in all cases, this is partly attributable to the metric itself and how it works, as the presence of shared or overlapping frequency structures among different modes can elevate spectral coherence despite the signals being distinct in content.


\subsection{Proof of Concept - Acoustic Sensing}

\begin{figure}[h!]
    \centering
    \includegraphics[width=\linewidth]{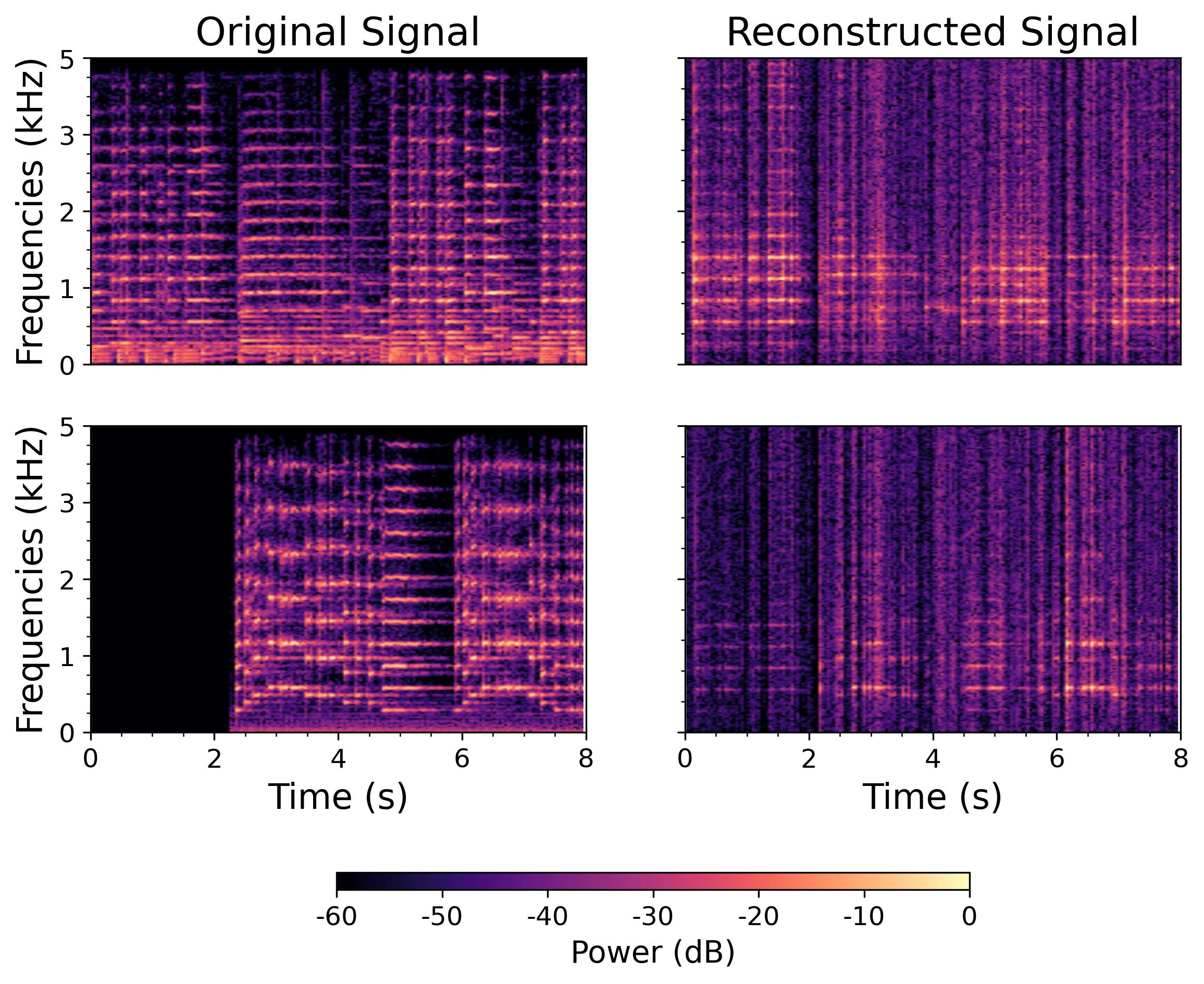}
    \caption{Original signal played on the two phones versus the signal reconstructed from the vibration of the boxes. The top spectrograms correspond to a trumpet cover of ABBA's "Dancing Queen," while the bottom plots show a standard trumpet sound effect.}
    \label{fig:results_box}
\end{figure}

Finally, to demonstrate the practical applicability of our approach for distributed acoustic sensing, we implemented a simplified proof-of-concept configuration. In this setup, we positioned the optical fiber on top of two plastic boxes, each containing a phone speaker directed at the center of the box from approximately 10\,cm distance. This configuration creates natural acoustic resonance chambers that couple sound to mechanical vibrations detectable by the fiber.
Following the same methodology established in previous experiments, we computed the OIM of the system using $f=600$\,Hz sinusoidal signals to determine the optimal interrogation modes. The frequency response of this configuration was primarily concentrated in the $f\in \left[ 400-1800 \right] $\,Hz range, with specific characteristics heavily influenced by the resonant properties (shape, size, and material) of the plastic boxes.
To test the performance of the system to complex acoustic inputs, we played single-instrument (trumpet) recordings through the speakers and reconstructed the signals using our tensor-based approach. Figure \ref{fig:results_box} presents a comparison between the spectrograms of the original audio inputs and their reconstructed counterparts. Again, the results presented demonstrate the ability of the system to maintain separation between the two acoustic sensing locations while preserving the key frequency structures and temporal dynamics of the trumpet recordings, with minimal cross-talk between channels.

Note that while the reconstruction quality in this proof-of-concept demonstration might be lower than the results presented in section \ref{sec:complex_audio}, with a more limited bandwidth, it still serves as a demonstrator for the capabilities of this interrogation scheme for practical multi-point acoustic sensing applications. The direct piezoelectric coupling used in the previous section provides superior signal fidelity but requires physical contact with the vibration source. Yet, this acoustic configuration demonstrates the feasibility of detecting airborne sound without direct mechanical coupling to the source. Indeed, future work could focus on enhancing the coupling efficiency between airborne acoustic signals and the optical fiber by exploring optimized resonance chamber designs, alternative materials with better acoustic properties, and improved structural configurations (e.g., fiber coils, mandrels) to extend the frequency response and increase sensitivity across different regions of the frequency spectrum.

\section{Concluding Remarks}\label{sec:discussion}

In this work, we report an innovative interrogation scheme for speckle-based optical fiber sensors using event-based sensors(EVS) together with a tensor-based methodology capable of separating and reconstructing simultaneous localized vibrations along a multimode optical fiber. First, focusing on the interrogation method with EVS, we showed that it allows to overcome the limitations of traditional speckle-based sensors, which are constrained by camera frame rates and dynamic range. By capturing only intensity variations with high temporal resolution in an asynchronous manner, the EVS results in high-speed speckle variation detections with low data transmission footprint, allowing for the reconstruction of frequencies covering the whole audible range and possibly entering the ultrasonic domain. Besides, its large dynamic range makes it an ideal sensor to interrogate speckle dynamics, overcoming under- or overexposure issues.

On the other hand, focusing on the introduced methodology for the determination of the optimal interrogation modes for analyzing speckle patterns, we showed that it allows both signal reconstruction and separation by modeling the response through data-driven calibration processes that account for non-ideal behavior of the system. Indeed, the data-driven procedure effectively creates specialized filters that maximize sensitivity to perturbations at specific locations while minimizing cross-talk, enabling truly distributed sensing along a single multimode fiber.

Our experimental results demonstrate that the proposed methodology successfully overcomes key challenges in speckle-based sensing. The frequency response characterization confirms robust operation across the entire audible spectrum, with only minor sensitivity reductions at certain frequencies. The successful reconstruction of complex audio signals with minimal cross-talk further validates our approach for real-world applications. Furthermore, the proof-of-concept acoustic sensing demonstration highlights the versatility of this methodology beyond laboratory conditions. While the acoustic coupling efficiency in this preliminary implementation was limited compared to direct piezoelectric actuation, it achieved satisfactory signal reconstruction for frequencies between 400-1800\,Hz.

The demonstrated ability to distinguish between multiple simultaneous perturbations enables applications ranging from structural health monitoring to acoustic surveillance. At the same time, the high temporal resolution provided by event-based detection makes this approach ideal for real-time monitoring of rapidly changing signals. In conclusion, our tensor-driven interrogation scheme leveraging disorder-enhanced sensitivity in multimode fibers offers unprecedented temporal resolution and spatial selectivity by transforming the inherent complexity of speckle patterns from a traditional limitation into a powerful sensing capability.

Finally, putting into a broader perspective besides acoustics, this approach also opens opportunities for real-time structural monitoring, soft robotics, and biomedical applications, where rapid and precise detection of mechanical perturbations is crucial \cite{jiang2023fiber, galloway2019fiber, garcia2015optical, ye2014structural, tosi2018fiber, roriz2013review}. Beyond these domains, the combination of asynchronous sensing and data-driven interrogation may be further explored in multiple directions, including edge processing to further reduce data rates, long-term stability strategies (thermal compensation and drift-aware recalibration), scaling to larger numbers of interrogation modes, and even hybrid architectures that fuse speckle interrogation with distributed acoustic sensing to marry fine temporal resolution with extended spatial coverage, offering a complementary and practical route to next-generation distributed and quasi-distributed fiber sensing.

\section{Acknowledgments}

Tomás Lopes and Joana Teixeira acknowledge the support of the Foundation for Science and Technology (FCT), Portugal, through Grants 2024.01830.BD and 2024.00426.BD, respectively.

\bibliographystyle{unsrt} 
\bibliography{sn-bibliography}

\begin{thebibliography}{10}

\bibitem{rotter2017light}
Stefan Rotter and Sylvain Gigan.
\newblock Light fields in complex media: Mesoscopic scattering meets wave control.
\newblock {\em Reviews of Modern Physics}, 89(1):015005, 2017.

\bibitem{hsu2017correlation}
Chia~Wei Hsu, Seng~Fatt Liew, Arthur Goetschy, Hui Cao, and A~Douglas~Stone.
\newblock Correlation-enhanced control of wave focusing in disordered media.
\newblock {\em Nature Physics}, 13(5):497--502, 2017.

\bibitem{byrnes2020universal}
Niall Byrnes and Matthew~R Foreman.
\newblock Universal bounds for imaging in scattering media.
\newblock {\em New Journal of Physics}, 22(8):083023, 2020.

\bibitem{popoff2010measuring}
S{\'e}bastien~M Popoff, Geoffroy Lerosey, R{\'e}mi Carminati, Mathias Fink, Albert~Claude Boccara, and Sylvain Gigan.
\newblock Measuring the transmission matrix in optics: An approach to the study and control of light propagation in disordered media.
\newblock {\em Physical review letters}, 104(10):100601, 2010.

\bibitem{gigan2022roadmap}
Sylvain Gigan, Ori Katz, Hilton~B De~Aguiar, Esben~Ravn Andresen, Alexandre Aubry, Jacopo Bertolotti, Emmanuel Bossy, Dorian Bouchet, Joshua Brake, Sophie Brasselet, et~al.
\newblock Roadmap on wavefront shaping and deep imaging in complex media.
\newblock {\em Journal of Physics: Photonics}, 4(4):042501, 2022.

\bibitem{gutierrez2024reaching}
Rodrigo Guti{\'e}rrez-Cuevas, Dorian Bouchet, Julien de~Rosny, and S{\'e}bastien~M Popoff.
\newblock Reaching the precision limit with tensor-based wavefront shaping.
\newblock {\em Nature Communications}, 15(1):6319, 2024.

\bibitem{bouchet2021maximum}
Dorian Bouchet, Stefan Rotter, and Allard~P Mosk.
\newblock Maximum information states for coherent scattering measurements.
\newblock {\em Nature Physics}, 17(5):564--568, 2021.

\bibitem{murray2019speckle}
Matthew~J Murray, Allen Davis, Clay Kirkendall, and Brandon Redding.
\newblock Speckle-based strain sensing in multimode fiber.
\newblock {\em Optics express}, 27(20):28494--28506, 2019.

\bibitem{cuevas2018machine}
Alberto~Rodr{\'\i}guez Cuevas, Marco Fontana, Luis Rodriguez-Cobo, Mauro Lomer, and Jos{\'e}~Miguel L{\'o}pez-Higuera.
\newblock Machine learning for turning optical fiber specklegram sensor into a spatially-resolved sensing system. proof of concept.
\newblock {\em Journal of Lightwave Technology}, 36(17):3733--3738, 2018.

\bibitem{gao2023spatially}
Han Gao and Haifeng Hu.
\newblock Spatially-resolved bending recognition based on a learning-empowered fiber specklegram sensor.
\newblock {\em Optics Express}, 31(5):7671--7683, 2023.

\bibitem{spillman1989statistical}
WB~Spillman~Jr, BR~Kline, LB~Maurice, and PL~Fuhr.
\newblock Statistical-mode sensor for fiber optic vibration sensing uses.
\newblock {\em Applied optics}, 28(15):3166--3176, 1989.

\bibitem{wu1991sensing}
Shudong Wu, Shizhou Yin, and Francis~TS Yu.
\newblock Sensing with fiber specklegrams.
\newblock {\em Applied optics}, 30(31):4468--4470, 1991.

\bibitem{okamoto1988multimode}
Takayuki Okamoto and Ichirou Yamaguchi.
\newblock Multimode fiber-optic mach-zehnder interferometer and its use in temperature measurement.
\newblock {\em Applied optics}, 27(15):3085--3087, 1988.

\bibitem{wang2017speckle}
Jiao-Jiao Wang, Shao-cheng Yan, and Fei Xu.
\newblock Speckle-based fiber sensor for temperature measurement.
\newblock In {\em 2017 16th International Conference on Optical Communications and Networks (ICOCN)}, pages 1--3. IEEE, 2017.

\bibitem{reja2024multimode}
Mohammad~Istiaque Reja, Darcy~L Smith, Linh~Viet Nguyen, Heike Ebendorff-Heidepriem, and Stephen~C Warren-Smith.
\newblock Multimode optical fiber specklegram pressure sensor using deep learning.
\newblock {\em IEEE Transactions on Instrumentation and Measurement}, 73:1--10, 2024.

\bibitem{yu1993submicrometer}
Francis~TS Yu, Meiyuan Wen, Shizhuo Yin, and Chii-Maw Uang.
\newblock Submicrometer displacement sensing using inner-product multimode fiber speckle fields.
\newblock {\em Applied optics}, 32(25):4685--4689, 1993.

\bibitem{gupta2008qualifying}
BHASKER Gupta, HN~Bhargaw, and HK~Sardana.
\newblock Qualifying fibre optic temperature sensor using speckle metrology.
\newblock {\em Int. Inf. Technol. Knowl. Manage}, 1(2):337--350, 2008.

\bibitem{lujo2008fiber}
Ivan Lujo, Pavo Klokoc, Tin Komljenovic, Marko Bosiljevac, and Zvonimir Sipus.
\newblock Fiber-optic vibration sensor based on multimode fiber.
\newblock {\em Radioengineering}, 17(2):93--97, 2008.

\bibitem{fujiwara2022optical}
Eric Fujiwara and Thiago~Destri Cabral.
\newblock Optical fiber specklegram sensor for multi-point curvature measurements.
\newblock {\em Applied Optics}, 61(23):6787--6794, 2022.

\bibitem{fujiwara2019optical}
Eric Fujiwara, Luiz~Evaristo da~Silva, Thiago~Destri Cabral, Hugo~Eug{\^e}nio de~Freitas, Yu~Tzu Wu, and Cristiano Monteiro de~Barros Cordeiro.
\newblock Optical fiber specklegram chemical sensor based on a concatenated multimode fiber structure.
\newblock {\em Journal of Lightwave Technology}, 37(19):5041--5047, 2019.

\bibitem{efendioglu2017review}
Hasan~Seckin Efendioglu.
\newblock A review of fiber-optic modal modulated sensors: Specklegram and modal power distribution sensing.
\newblock {\em IEEE Sensors Journal}, 17(7):2055--2064, 2017.

\bibitem{wei2021neural}
Menglong Wei, Gang Tang, Jie Liu, Luyang Zhu, Junyi Liu, Cong Huang, Jingxing Zhang, Lei Shen, and Siyuan Yu.
\newblock Neural network based perturbation-location fiber specklegram sensing system towards applications with limited number of training samples.
\newblock {\em Journal of Lightwave Technology}, 39(19):6315--6326, 2021.

\bibitem{lenero2018applications}
Juan~A Le{\~n}ero-Bardallo, Ricardo Carmona-Gal{\'a}n, and Angel Rodr{\'\i}guez-V{\'a}zquez.
\newblock Applications of event-based image sensors—review and analysis.
\newblock {\em International Journal of Circuit Theory and Applications}, 46(9):1620--1630, 2018.

\bibitem{gallego2020event}
Guillermo Gallego, Tobi Delbr{\"u}ck, Garrick Orchard, Chiara Bartolozzi, Brian Taba, Andrea Censi, Stefan Leutenegger, Andrew~J Davison, J{\"o}rg Conradt, Kostas Daniilidis, et~al.
\newblock Event-based vision: A survey.
\newblock {\em IEEE transactions on pattern analysis and machine intelligence}, 44(1):154--180, 2020.

\bibitem{ploschner2015seeing}
Martin Pl{\"o}schner, Tom{\'a}{\v{s}} Tyc, and Tom{\'a}{\v{s}} {\v{C}}i{\v{z}}m{\'a}r.
\newblock Seeing through chaos in multimode fibres.
\newblock {\em Nature photonics}, 9(8):529--535, 2015.

\bibitem{bianchi2012multi}
Silvio Bianchi and Roberto Di~Leonardo.
\newblock A multi-mode fiber probe for holographic micromanipulation and microscopy.
\newblock {\em Lab on a Chip}, 12(3):635--639, 2012.

\bibitem{tiwari1999nonlinear}
BB~Tiwari, Vijay Prakash, Vibha Tripathi, and N~Malaviya.
\newblock Nonlinear effects in optical fiber transmission system.
\newblock {\em IETE Technical Review}, 16(5-6):461--479, 1999.

\bibitem{zhang2011photonic}
Min Zhang, Xiaohong Ma, Liwei Wang, Shurong Lai, Hongpu Zhou, Huafeng Zhao, and Yanbiao Liao.
\newblock Photonic sensors review progress of optical fiber sensors and its application in harsh environment.
\newblock {\em Photonic Sensors}, 1:84--89, 2011.

\bibitem{zhang2006fiber}
Zhijun Zhang and Farhad Ansari.
\newblock Fiber-optic laser speckle-intensity crack sensor for embedment in concrete.
\newblock {\em Sensors and Actuators A: Physical}, 126(1):107--111, 2006.

\bibitem{Lopes_Harnessing_Disorder_Tensor-Driven}
Tomás Lopes, Joana Teixeira, Vicente V.~Rocha, Tiago D.~Ferreira, Catarina S.~Monteiro, Pedro A.S.~Jorge, and Nuno A.~Silva.
\newblock {Harnessing Disorder: Tensor-Driven Distributed Acoustic Sensing in Optical Fibers via High-Frequency Event-Based Speckle Interrogation}.

\bibitem{jiang2023fiber}
Qi~Jiang, Jihua Li, and Danish Masood.
\newblock Fiber-optic-based force and shape sensing in surgical robots: a review.
\newblock {\em Sensor Review}, 43(2):52--71, 2023.

\bibitem{galloway2019fiber}
Kevin~C Galloway, Yue Chen, Emily Templeton, Brian Rife, Isuru~S Godage, and Eric~J Barth.
\newblock Fiber optic shape sensing for soft robotics.
\newblock {\em Soft robotics}, 6(5):671--684, 2019.

\bibitem{garcia2015optical}
Iker Garc{\'\i}a, Joseba Zubia, Gaizka Durana, Gotzon Aldabaldetreku, Mar{\'\i}a~Asunci{\'o}n Illarramendi, and Joel Villatoro.
\newblock Optical fiber sensors for aircraft structural health monitoring.
\newblock {\em Sensors}, 15(7):15494--15519, 2015.

\bibitem{ye2014structural}
XW~Ye, YH~Su, and JP~Han.
\newblock Structural health monitoring of civil infrastructure using optical fiber sensing technology: A comprehensive review.
\newblock {\em The Scientific World Journal}, 2014(1):652329, 2014.

\bibitem{tosi2018fiber}
Daniele Tosi, Sven Poeggel, Iulian Iordachita, and Emiliano Schena.
\newblock Fiber optic sensors for biomedical applications.
\newblock In {\em Opto-mechanical fiber optic sensors}, pages 301--333. Elsevier, Amsterdam, Netherlands, 2018.

\bibitem{roriz2013review}
Paulo Roriz, Orlando Fraz{\~a}o, Ant{\'o}nio~B Lobo-Ribeiro, Jos{\'e}~L Santos, and Jos{\'e}~A Sim{\~o}es.
\newblock Review of fiber-optic pressure sensors for biomedical and biomechanical applications.
\newblock {\em Journal of biomedical optics}, 18(5):050903--050903, 2013.

\end{thebibliography}

\end{document}